# Solar Forcing on the Changing Climate

**K. M. Hiremath**

*Indian Institute of Astrophysics, Bangalore-560034, India*
hiremath@iiap.res.in

**Abstract.** The influence of solar cycle and activity phenomena on the two climatic variabilities such as the rainfall and the surface temperature of the Indian subcontinent are critically examined. It is concluded from this study that the sun indeed strongly influences both of these climatic variabilities and the sun's influence simply cannot be ignored.

## 1. Introduction

From the dawn of the civilization, the sun is revered and held as an awe inspiring celestial object inthe sky. In the world, there are many stories and poems woven around the god sun in different folklores and magnificent architectures are dedicated to the mighty sun. The flora and fauna on the earth mainly depend on the sun for their survival. To sustain the life on the earth, energy is derived from the sun.

Does also the sun sustain and change the planetary climates in general and the earth's climate in particular? Do physical parameters of the earth's surface or atmosphere vary in consonance with the solar cycle and activity phenomena? Yes, the ample evidences from the scientific literature (Reid 1999; Shine 2000; Unno and Shibahashi 2003; Hiremath and Mandi 2004 and references therein; Sakurai, Rusin and Minarovjech 2004; Georgieva, et. al. 2005a; Georgieva and Kirov 2006; Soon 2005 and references there in; Valev 2006; Hiremath 2006a; Hiremath 2006b; Haigh 2007; Badruddin, Singh and Singh 2006; Perry 2007; Feymann 2007; Tiwari and Ramesh 2007 and references therein; Scafetta and West 2008; Soon 2009) show that the sun indeed influences the earth's climate and environment. Analysis of vast stretch in time of the paleoclimatic records (Beer, Tobias and Weiss 1998; Muscheler et. al., 2007) show that the sun's activity is imprinted in the global temperature (Georgieva, Bianchi and Kirov 2005 and references therein; Valev 2006) and precipitation variabilities. Owing to the proximity of the earth to the sun, one can not neglect sun's influence on the technological and biological systems of the human society (Babayev 2006; Babayev et. al. 2007). As the human society is advancing in space technology and moving in future to other planets, one should know in advance the space weather effects mainly due to the sun.

The earth's global temperature and precipitation such as rainfall - two vital parameters of the climatic system - need to be studied carefully. These two important physical parameters of the global climate affect the human society at large. We have to learn from the past historical records. Some of the scientific evidences show that drastic and catastrophic changes in the climate due to either temperature or precipitation patterns (like the floods and droughts) lead to the end of civilizations (for example the Mayan culture, maybe Harappa and Mohenjo-Daro) on our planet. Until the advent of industrialization, it was believed that the variation of sun's energy output strongly influenced the earth's environment and climate. However, after the industrial era, the anthropogenic influences on climate are dominating compared to the solar influence. The increase in the earth's global surface temperature is recently believed to be attributed especially to the phenomenal increase of emission of green house gases like carbon dioxide mainly contributed by the human beings. As the climate of the Indian subcontinent is strongly related to the global climatic variations, from the correlative analysis of two climatic records of temperature and the Monsoon rainfall of the Indian subcontinent it is argued in the present study that the sun's influence cannot be ignored and should not be underestimated.

## 2. Brief Introduction of the Sun

The important physical parameters of the sun are: (i) mass - $2 \times 10^{30}$ Kg, (ii) radius - $7 \times 10^{8}$ meter, (iii) mean density-1409 Kg/m$^{-3}$, (iv) temperature at the photosphere - 5780 $^{o}$K, and (v) the total amount of energy emitted by the sun (i.e., luminosity) measured at one astronomical unit (i.e., at the distance between the earth and the sun) - $3.9 \times 10^{26}$ Joule/sec.

When one considers the cross section of the sun, parallel to it's rotation axis, based on the dynamical and physical properties, the sun's interior can be classified into three distinct zones: (i) the *radiative core* where the energy is generated by the nuclear fusion of hydrogen atoms and is transferred by the radiation, (ii) the *convection zone* where the energy is transferred from the base of the convection zone to the surface by the convection of the plasma and, (iii) the *photosphere* where the energy is radiated to the space. Above the photosphere, the sun's atmosphere consists of the *chromosphere* and the *corona*. The temperature increases from the layer of the photosphere to million degree Kelvin in the corona.

If the sun were static in time with constant output of energy, the planetary environments in general and the earth in particular would have received the constant output of energy at their surfaces. However, the sun's energy output is variable (~ 0.1% in the total irradiance and ~ 10-20% in the UV 200-300 nm) due to spatial and temporal variability of the sun's large scale magnetic field structure, dynamics and flow of mass (both the neutral and charged particles) from the sun. The most outstanding activity of the sun is the *sunspots* - cool and dark features compared to the ambient medium - on the sun's surface that modulate the sun's irradiance and the galactic cosmic rays that enter in the planetary environments. The *flares* that are associated with the sunspots (Hiremath 2006c) release vast amount of energy (~ $10^{20}$ - $10^{25}$ Joule) within a short span of time. The sun also ejects sporadically a mass of plasma (~ $10^{12}$ Kg) to the space which is called *coronal mass ejection*.

There is also a continuous flow of wind (~ $10^{31}$ charged particles per second or 6-7 billion tons per hour) from the sun towards the space called *solar wind*. The sun's activity varies on time scales of few minutes to months, years to decades and perhaps more than centuries. The 5 minute global oscillations are due to pressure perturbations in the solar interior. The solar different activity indices vary on the time scales of ~ 27 days due to solar rotation and ~ 150 days due to the flares. The ~ 1.3 year periodicity is predominant in different solar activity indices. The next prominent and ubiquitous *viz.*, 11 year solar cycle periodicity that is found not only in the present day sun's activity indices but also in the past evolutionary history indices that are derived from the solar proxies (such as $C^{14}$ and $Be^{10}$).

The sun can influence the earth's climate and environment in two ways: (i) by direct influence (Kilifarsha 2006) on the atmosphere (especially the stratosphere and troposphere), the surface and the ocean, and (ii) by indirect influence through the galactic cosmic rays (Bucharova and Velinov 2006). Directly the sun can influence the earth's surface, atmosphere and oceans by its electromagnetic radiation in all wavelengths. Indirectly the sun mediates the amplitude variation of the galactic cosmic rays that deeply penetrate and ionize the earth's atmosphere leading to the formation of rain drops.

# 3. Data, Results and Conclusions

In the present study, for the period of 1871-2005, we consider the data of the sunspots and irradiance variations. As for the earth's climate, both the temperature and Indian rainfall variabilities are considered from the website "http://www.tropmet.res.in" maintained by Rupa Kumar and his colleagues. The Indian rainfall variability has similar characteristics and association with the regional or global circulation parameters. In the present study, we use the seasonal and annual values (averaged for the periods of 12 months). For the years 1871-2005, the rainfall data (in mm) are available in monthly and annual series. The sunspot data are obtained from the website "http://www.ngdc.noaa.gov/STP/SOLAR" and the two data set of solar (total and UV in the wavelength range of 200-295 nm) irradiance variabilities are kindly provided by Dr. Lean (Naval Observatory, USA). In Figure 1, the annual sunspot and the Indian rainfall variabilities are illustrated. The solar irradiance (dotted) and sunspot (continuous) variabilities are presented in Figure 2.

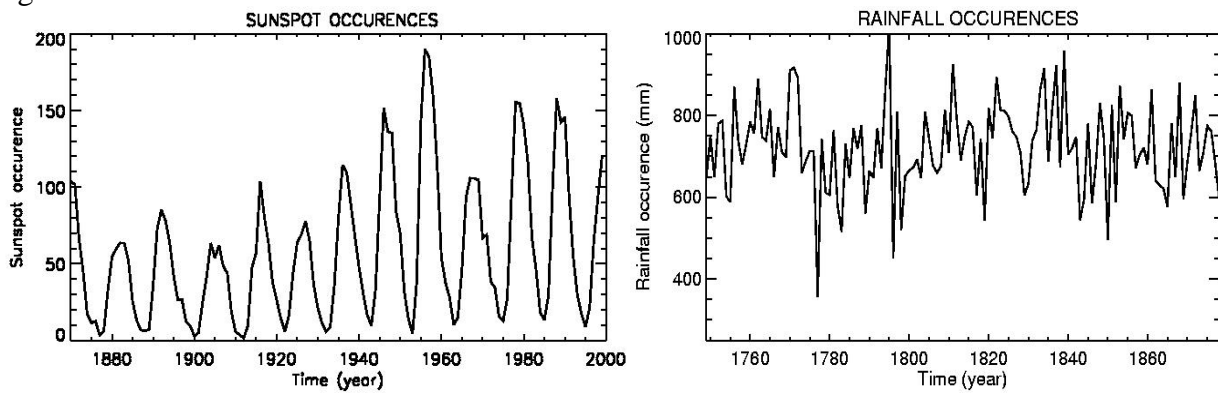

**Fig 1 .** (a) *Temporal variation of the occurrence of the annual sunspot numbers. (b) Temporal variation of occurrence of the annual Indian rainfall.*

In order to assess the influence of the sun on the earth's climate, we perform a correlative analysis. For confirming whether the solar influence on the rainfall variability is different in different seasons and following our previous study (Hiremath and Mandi 2004), we classify the data set into four seasons: (i) spring (March-May) rainfall, (ii) south west monsoon (June-Sept) rainfall, (iii) north east monsoon (Oct-Dec) rainfall, and (iv) winter rainfall (Jan & Feb).

## 3.1 Correlative analysis of the solar and rainfall variabilities

In the previous study (Hiremath and Mandi 2004), we performed a correlative analysis of the sunspots and Indian rainfall variabilities for the annual as well as for the seasonal data sets. In the present study, we extend the data up to the year 2005 and perform the same correlative analysis. As in the previous study, for the seasonal sunspot and Indian rainfall variabilities, we obtained a moderate positive correlation (correlation coefficient ~ 30%) for all the data set and the results are illustrated in Fig 3. We also performed the same correlative analysis for the solar annual total irradiance and the Indian rainfall variabilities. The correlation improves slightly, but not better. In Fig 4, the annual means of total irradiance and the rainfall variabilities are presented. However, when we perform a correlative analysis for the UV irradiance in the wavelength range of 200-295 nm, especially for MAM and southwest monsoon (JJAS) rainfall veriabilities, we get a very good correlation (~ 60%) and these results are illustrated in Fig 5.

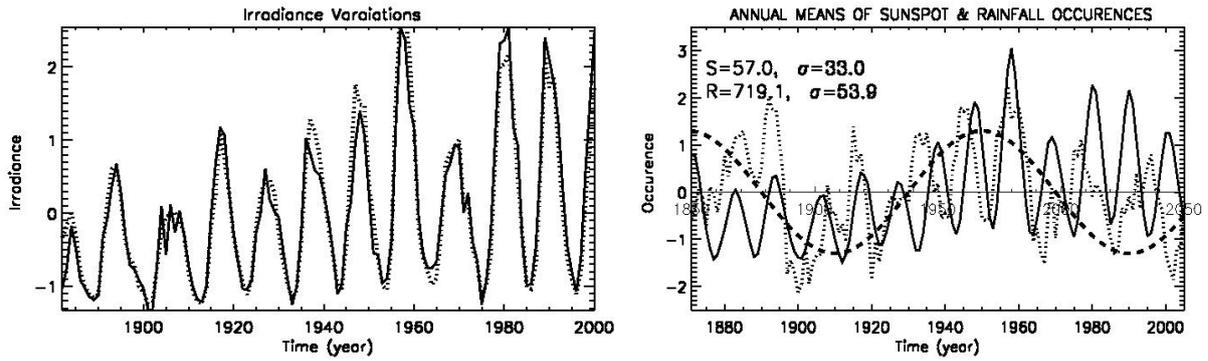

Fig 2. *The deviation from the mean and normalized to their respective standard deviations of the annual solar and Indian rainfall occurrence variabilities. (a) In the left figure, the continuous line is the occurrence of sunspot number and the dotted line is irradiance variations, (b) In the right figure, the continuous line is sunspot occurrence variability and the dotted line is the annual Indian rainfall variability. The thick dashed curve is a long-term (~ 80-100 years) variation of the annual Indian rainfall variability. S and R are the means of the sunspot and rainfall occurrence variabilities. σ is the standard deviation from the mean of both the variabilities.*

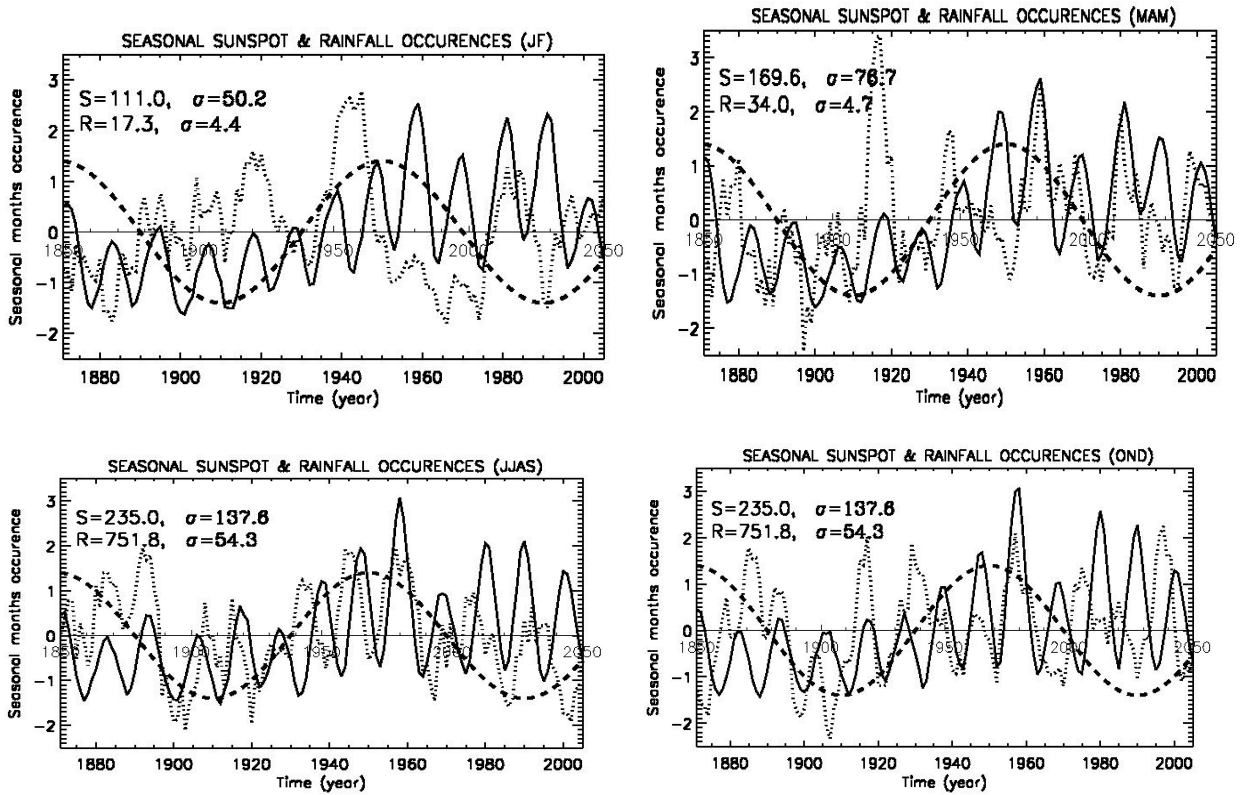

Fig 3. *The deviation from the mean and normalized to their respective standard deviations of the seasonal sunspot (continuous lines) and Indian rainfall occurrence (dotted lines) variabilities. In all the above four figures, S is the mean of the sunspot activity, R is the mean of the rainfall variability and σ is the standard deviation from the mean for both the variabilities. The thick dashed curve is a long-term (~ 80-100 years) variation of the annual Indian rainfall variability.*

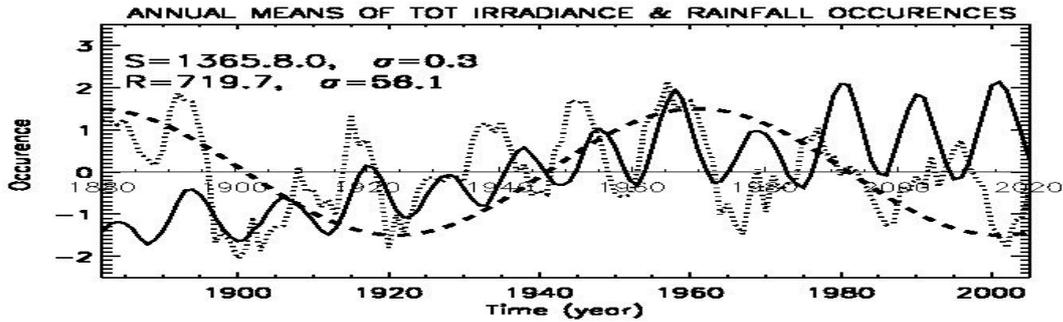

Fig 4. *The deviation from the mean and normalized to their respective standard deviations of the annual total irradiance (continuous line) and Indian Monsoon rainfall (dotted line) variabilities. S is the mean of the irradiance activity, R is the mean of the rainfall variability and σ is the standard deviation from the mean. The thick dashed line is a long-term (~ 80-100 years) variation of the annual Indian rainfall variability.*

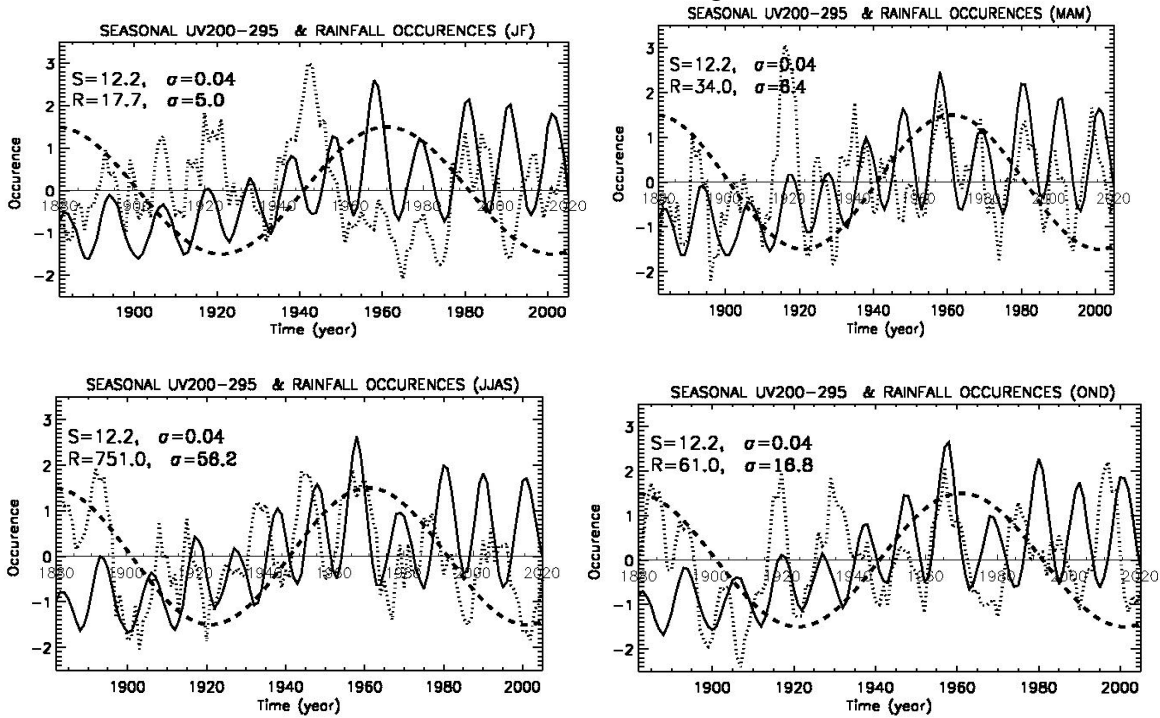

Fig 5. *The deviation from the mean and normalized to their respective standard deviations of the seasonal irradiance (continuous line) in the wavelength range of 200-295 nm and the Indian Monsoon rainfall (dotted line) variabilities. S is the mean of the irradiance activity, R is the mean of the rainfall variability and σ is the standard deviation from the mean for both the variabilities. The thick dashed line is a long-term (~ 80-100 years) variation of the annual Indian rainfall variability.*

Although the visual inspection of Figures 2(b)-5 shows a good correlation between the two solar and rainfall variabilities, computed and quantitative information such as the correlation coefficient is found to be at most of ~ 60% . Moreover from 1960 onwards the rainfall variability is out of phase with the solar variabilities. This possibly suggests that the variations of the Indian annual rainfall variability may not be due to annual variation of the sunspot and solar irradiance variabilities. Hence, one should be cautious enough to claim that solar activity overwhelmingly influences the Indian rainfall. One should not also be under impression that sunspot and irradiance variabilities are only the main solar activities. One should also check with other solar activity phenomena such as solar wind, coronal holes (Soon et. al. 2000; Georgieva and Kirov 2006) and, coronal mass ejections. Other important reason for the low correlation could be due to not removing the unknown long term trends in both the rainfall and solar variabilities.

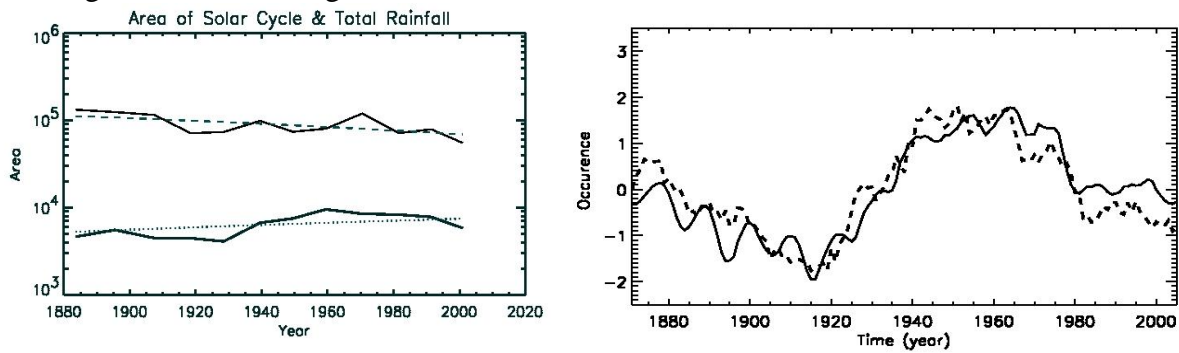

**Fig 6.** (a) *The left figure represents the long term variation of the areas of sunspot (increasing trend) and the rainfall (decreasing trend) variabilities. The dotted and dashed lines are the linear least square fit to the respective data,(b) the right figure (by removing long term decreasing trend in the rainfall and increasing trend in the solar activity)illustrates the Indian Monsoon rainfall and sunspot activities. The continuous line is the occurrence of sunspot activity and dashed line is the rainfall variability.*

For the results presented in Fig 2(b)-Fig 5, we removed the annual trends by subtracting the respective means and normalized with respect to their standard deviations, and came to the conclusion that we did not get very good correlation. Let us probe further whether we get a good correlation in case we remove the decadal variations. In order to remove the decadal variations, following Hiremath and Mandi (2004) and for different periods of the solar cycle, first we compute areas of the solar cycles and then the resulting areas are subjected to the linear least square fit. One can notice from Fig 6(a) (continuous line overplotted on the dotted line) that during the period of 1875-2005, there is an increasing trend of the solar activity. The known increasing trend is removed from the smoothened sunspot variability (by taking moving averages of ten years or more from the annual data).  Similar procedure for detrending the data is applied to the rainfall variability. From Fig 6(a) (continuous line overplotted on the dashed line), one can notice that there is a long-term decreasing trend in the annual rainfall variability. This long-term decreasing trend is most evident if we extend the present day rainfall variability to the past rainfall variability (obtained by the proxy indices) over hundreds of years back. Both the long-term detrended data are presented in Fig 6(b).When we compute the correlation coefficient of the smoothened long-term detrended data of both rainfall and sunspot data, we get a very good correlation of ~ 80-90% (with high significance) depending upon moving average intervals of 10-22 years. Thus we can safely conclude that, on long-term scale (when moving averages of 10 or more years are considered), sun indeed influences the Indian rainfall activity. Owing to the strong solar influence on the Indian Monsoon variability and the predictive capability of future solar activity cycles (Hiremath 2008), one can predict the long-term Indian rainfall variability by 10-100 years in advance.

## 3.2 Correlative analysis of the solar and temperature variabilities

In order to assess the solar influence on other climatic variability of the Indian subcontinent, we consider the available (1901-2003) maximum, minimum and difference between the maximum and minimum (that represents the earth's albedo) temperature variabilities. As described in the previous section, we computed the deviation from the mean and then normalized with respect to their standard deviations.

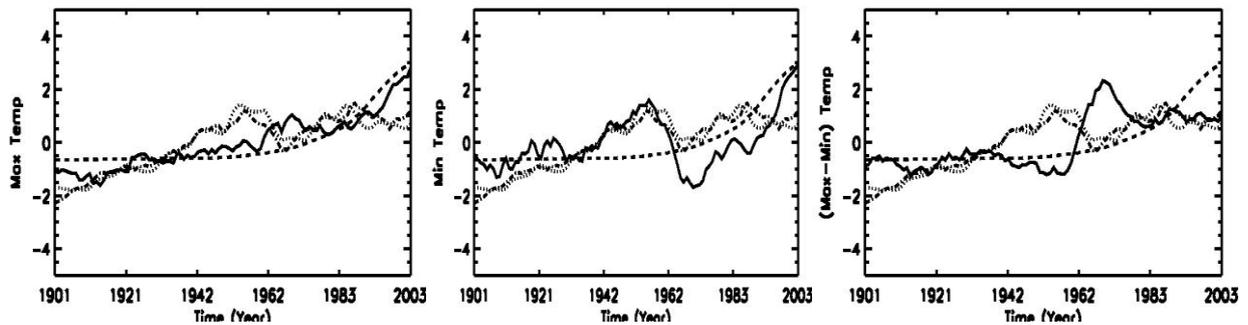

*Fig 7. The deviation from the mean and normalized to the respective standard deviations of the temperature of the Indian subcontinent, Asian atmospheric concentration of carbon dioxide and solar activity variabilities. The continuous line is the surface temperature of the Indian subcontinent, the dashed line is the concentration of atmospheric carbon dioxide, the dotted line is the solar UV irradiance and the dash-dotted line is the geomagnetic aa index. All the three figures (fig 7(a), fig 7(b) and fig 7(c)) illustrate the maximum, minimum and the difference of maximum and minimum (maximum-minimum) temperature variabilities over plotted with the solar activity indices.*

As the increase in the global warming is believed to be related with increase in the concentration of the atmospheric carbon dioxide, for sake of the comparison, the concentration of the atmospheric carbon dioxide (http://cdiac.esd.oml.gov) of the Asian continent is considered. As for the solar activity indices, we consider not only the sunspot and UV irradiance data but also the geomagnetic aa index - supposed to be the indicator of solar activity (Georgieva and Kirov 2006). Consideration of the geomagnetic aa index for this correlative analysis basically also is influenced by the previous similar studies (Goergieva, Bianchi and Kirov 2005; Georgieva, Kirov and Bianchi 2005c). In Fig. 7, we present the surface temperature of the Indian subcontinent and the atmospheric carbon dioxide of the Asian continent together with different solar activity indices. One important conclusion from this illustration is that, similar to the global warming, Indian sub continent is also warming substantially especially after 1980 onwards. One can also find a similar trend of the concentration of the atmospheric carbon dioxide of the Asian continent. Although visual inspection and computation of the correlation coefficient appear to yield a very good correlation between the temperature and carbon dioxide variabilities, one cannot explain the annual and decadal variations of the temperature variability. This is because the increase in the atmospheric carbon dioxide variability, especially after 1980, is very smooth compared to either temperature or solar activity variabilities.

Following are other important results from this correlative study between the temperature of the Indian subcontinent and the solar activity indices: (i) after year 1980 onwards, although there is a strong correlation between the variation of maximum temperature and the variation of concentration of carbon dioxide in the atmosphere of the Asian continent, the records of variation of minimum temperature and difference between maximum and minimum temperature records (that represents the earth's albedo) vary strongly with the solar activity, (ii) the maximum temperature variability also yields a strong correlation with the variation of solar activity with a lag of six year, and (iii) there is a linear increasing trend or pattern in the variation of the concentration of carbon dioxide in the century scale data, (iv) on the other hand, in addition to the increasing linear trend, there are superposed annual and decadal variations in the surface temperature variability whose amplitude variations are almost similar to the amplitude variations of the solar activity. From all these four results, one can safely conclude that the solar activity also strongly influences the surface temperature variability of the Indian subcontinent. The overall conclusion from the previous and present sections is that *there is a strong forcing on the changing climate of the Indian subcontinent.*

## 4. Conclusions

In order to assess the influence of solar activity on the climate of the Indian sub continent, two variabilities data such as rainfall and temperature are considered for the present analysis. Correlative analysis between the solar activity indices and two climate variabilities of the Indian subcontinent suggests that there is a strong solar forcing on the changing climate of the Indian subcontinent and it can not be simply ignored.

## 5. Acknowledgements

Author is grateful to Dr. Katya Georgieva, Dr. Haubold and SOC for providing the full financial support for attending the conference.

## References

Beer, J., Tobias and Weiss, W., 1998, Sol Phys., vol 181, p. 237
Babayev, E. S., 2006, "Space weather influence on Technological, biological and ecological system", Sun and Geosphere, vol 1 (1), p. 12-16
Babayev, E. S., Allahaverdiyava, A. A., Mustafa, F. R and Shusterev, P. N., 2007, "An influence of Changes of heliogeophysical conditions on biological systems : some results of studies conducted in the Azerbajan National Academy of Science., Sun and Geosphere, vol 2., p. 48-52
Badruddin, Singh, Y. P and Singh, M., 2006, "Does solar variability affect Indian (Tropical) weather and climate? : An assessment", in the proceedings of ILWS workshop, Goa, Editors : N. Gopalswamy and A. Bhattacharyya., p. 444
Buchvarova, M and Velinov, P. I. Y., 2006, "Cosmic rays and 11-year solar modulation", Sun and Geosphere, vol 1(1), p. 27-30
Feymann, J., 2007, "Has solar variability caused climatic change that affected human culture", Advances in Space Science, vol. 40, p. 1173
Georgieva, K., Bianchi, C and Kirov, B., 2005b, "Once again about global warming and solar activity", "Memorie dell Socirta Astronomica Italiana, vol. 76, p. 969
Georgieva, K., Kirov, B and Bianchi, C., 2005c, "Long-term variations in the correlation between solar activity and climate", "Memorie dell Socirta Astronomica Italiana, vol. 76, p. 969
Georgieva, K., Kirov, B., Javaraiah, J and Krasteva, R., 2005a, "Solar rotation and solar wind magnetospheric coupling", Planetary and Space science, vol. 53., p. 197-207
Georgieva, K and Kirov, B., 2006, "Solar activity and global warming revisited", "Sun and Geosphere", vol 1(1), p. 12-16

Haigh, D., 2007, "The sun and the Earth's climate", Living Reviews in Solar Phys, vol 4, p. 2

Hiremath, K. M and Mandi, P. I., 2004, "Influence of the solar activity on the Indian Monsoon rainfall", New Astronomy, vol 9, 651

Hiremath, K. M., 2006a, "The Influence of Solar Activity on the Rainfall Over India: Cycle to Cycle Variations", Journal of Astrophysics and Astronomy, vol 27, 367

Hiremath, K. M. 2006b, "Influence of solar activity on the rainfall over India", in the proceedings of ILWS workshop, Goa, Editors : N. Gopalswamy and A. Bhattacharyya., p. 178

Hiremath, K. M., 2006c., "The flares associated with the dynamics of the sunspots", Journal of Astrophysics and Astronomy, vol. 27 (no 2 and 3), p. 277-284

Hiremath, K. M., 2008, "Prediction of solar cycle 24 and beyond", Astrophysics and Space Science, vol. 314, p. 45-49

Kilifarska, N., 2006, "Solar variability and climate-UTLS amplification of solar signal", Sun and Geosphere, vol 1(1)

Muscheler, R ., et. al. 2007., Queternary Science Reviews, 26, 82

Perry, C. A., 2007, "Evidences for linkage between galactic cosmic rays and regional climate series", Advances in Space Res, vol 40, p. 353

Reid, G. C., 1999, "Solar Variability and it's implications for the human environment", Journ. Atmos. Solar. Terr. Phys, 61, p. 3

Scafetta, N and West, B. J., 2008, "Is climate sensitive to solar variability", Physics Today, March Issue, p. 50-51

Sakurai, T., Rusin, V and Minarovjech, M., 2004, "Solar-cycle variation of near-sun sky brightness observed with coronagraphs", Advances in Space Research, vol 34, 297

Shine, K. P., 2000, "Radiative forcing of climate change". Space Science Reviews, vol 94, p. 363

Soon, W., Baliunas, S., Posmentier, E. S. and Okeke, P., 2000, "Variations of solar coronal hole area and terrestrial lower tropospheric air temperature from 1979 to mid-1998: astronomical forcings of change in earth's climate? ", New Astronomy, vol 4, no. 8., p., 563

Soon, W., 2005, "Variable solar irradiance as a plausible agent for multidecadal variations in the Arctic-wide surface air temperature record of the past 130 yrs". Geophys. Res. Let., vol 32, 16, p., L16712

Soon, W., 2009, "Solar Arctic-mediated climate variation on multidecadal to centennial timescales: Empirical evidence, mechanistic explanation, and testable consequences", to appear in Physical Geography

Tiwari, M and Ramesh, R., 2007, "Variability of climate change in India", Current Science, vol 93, 477

Unno, W and Shibahashi, H., 2003, in "Stellar astrophysical fluid dynamics", Editors: M. J. Thompson and Dalsgaard, J. C., p. 411

Valev, D, 2006, "Statistical relationships between the surface air temperature anomalies and the solar and geomagnetic activity indices", Physics & Chemistry of the Earth, vol 31, p. 109